\documentstyle[prl,aps,epsf,multicol]{revtex}
\begin{document}
\begin{multicols}{2}


\noindent
{\large \bf Comment on 
``Quantum Decoherence in Disordered Mesoscopic Systems''}
\vspace{0.3cm}

In a recent letter\cite{Zaikin1}, Golubev and Zaikin (GZ) found
that  ``zero-point fluctuations of electrons''
contribute to the
dephasing rate $1/\tau_\varphi$ extracted from the magnetoresistance. 
As a result, $1/\tau_\varphi$  remains finite at zero temperature, $T$. 
GZ claimed that their results ``agree well with the
experimental data''.

We point out  that the GZ results
are 
{\em incompatible} with (i)  conventional perturbation theory of the
 effects of interaction on weak localization (WL), and (ii) with the 
 available experimental data.
More detailed criticism of Ref.~\cite{Zaikin1} 
can be found in Ref.~\cite{preprint}.

According to Ref.~\onlinecite{Zaikin1}, as $T\to 0$ in all dimensions
\begin{equation}
\frac{\hbar}{\tau_{\varphi}} = \frac{\hbar}{\tau\ g[L^\ast]}, \quad
L^\ast=\sqrt{D\tau},
\label{zaikin}
\end{equation}
where $\tau$
is the elastic time, $D$ is the diffusion constant, and
$g[L] \propto L^{d-2}$ is the  conductance [in units of
$e^2/(2\pi\hbar)$] of a sample of  size $L$. 

This result differs from the conventional one\cite{AAK,AA} 
\begin{equation}
\frac{\hbar}{\tau_{\varphi}} \simeq \frac{T}
{g\left[L^\ast
 \right]},\quad 
\label{tauphi}
L^\ast =
{\rm min}\left(\sqrt{D\tau_\varphi},
\sqrt{D\tau_H}\right),
\end{equation}
where $\tau_H$ 
is the scale due to the breaking of the time-reversal invariance
by the magnetic field, $H$~\cite{AA}.

The idea of the zero-$T$ dephasing
can be rejected using qualitative arguments 
(Secs. 2 and 3 of Ref.~\cite{preprint}). Here 
we provide  the result of the formal calculation.
The dephasing contribution for
$\tau_\varphi \gtrsim \tau_H$ can be found from
the expansion of the WL correction to the conductivity
\begin{equation}
\frac{\delta\sigma}{\sigma}\simeq -\frac{1}
{g\left(\sqrt{D\tau_H}\right)}+
\frac{1}
{g\left(\sqrt{D\tau_H}\right)}\frac{\tau_H}{\tau_\varphi}
+\dots,
\label{expansion}
\end{equation}
and the second term on the RHS appears
in the first-order
perturbation theory in the interaction propagator. The
calculation which takes into account {\em all the diagrams} of 
the order of $1/g^2$, 
(Secs. 4,5 of Ref.~\cite{preprint}), leads to
\end{multicols}
\widetext
\begin{eqnarray}
\delta\sigma_{I\times WL}&=&\frac{e^2}{\pi\hbar} 
\frac{e^2}{\hbar\sigma_1}
\left\{
D\tau_H
\left(\frac{T\tau_H}{4\hbar}\right)
\left[1+\zeta\left(\frac{1}{2}\right)
\sqrt{\frac{2\hbar}{\pi T\tau_H}}
\right]
+
\frac{\zeta\left(\frac{3}{2}\right)}{\pi}
\sqrt{\frac{ \hbar D^2\tau_H}{2\pi T}}
\right\}, \quad d = 1, \nonumber\\
\delta\sigma_{I\times WL}&=&
\frac{e^2}{2\pi^2\hbar} \frac{R_\Box e^2}{2\pi^2\hbar}
\left\{
 \frac{\pi T\tau_H}{\hbar}
\left[
\ln \left(\frac{T\tau_H}{\hbar}\right)
+1
\right]
+\frac{3}{2}\ln \left(\frac{\tau_H}{\tau}\right)
+{\cal O}\left[\ln \left(T\tau/\hbar\right)\right]
\right\}
,
\quad d=2
\label{results}
\end{eqnarray}
where $\sigma_1$ is the conductivity per unit length of a
one-dimensional conductor, $R_\Box$ is the sheet resistance of a
two-dimensional film, 
$\zeta (1/2) = -1.461\dots$, $\zeta (3/2) = 2.612\dots$. 
\vspace*{-0.2cm}
\begin{multicols}{2}
Comparison of Eqs.~(\ref{results}) with 
Eq.~(\ref{expansion}) shows that $\tau_\varphi$ is given by
Eq.~(\ref{tauphi}) rather than by Eq.~(\ref{zaikin}). 
The procedure of Ref.~\cite{Zaikin1} is nothing but a perturbative expansion.
Since it disagrees parametrically 
with the diagrammatic expansion already in the first order, it
is simply wrong.
The errors of Ref.~\cite{Zaikin1} stem  
from the uncontrollable procedure of the 
semiclassical averages; as a result, 
some contributions were lost (Sec.~6.1
of Ref.~\cite{preprint}).

The results of Ref.~\cite{Zaikin1} are in contradiction with the experiments.  
It is well known that the magnetoresistance in $2d$ and $3d$ systems
(quasi-$2d$ and $3d$ metal films, metal glasses, $3d$ doped
semiconductors, $2DEG$ in heterostructures, {\em etc.}) 
 depends substantially on the temperature.
Such a dependence is impossible according to Ref.~\cite{Zaikin1}. 
Indeed, for disordered metals with $\tau=10^{-16} \mbox{---}
10^{-14}s$, Eq.~(\ref{zaikin}) predicts 
a  $T$-independent dephasing rate  for any conceivable temperature. 
The experimental values of $\tau_\varphi$
exceed by far the estimates Eq.~(\ref{zaikin}): e.g., by $10^5$ for the
$3d$ Cu films\cite{AGZ} (for a more detailed comparison of the
experimental data on $\tau_\varphi$ with Eq.~(\ref{zaikin}) 
see Sec.~6.2 of Ref.~\cite{preprint}).  
The statement\cite{Zaikin1} that the
interactions preclude the crossover into the
insulating regime in low-dimensional conductors, is also at odds with
 experiment. The weak-to-strong localization crossover has been
observed for both $1d$ and $2d$ cases (see
e.g. Refs.~\cite{hsu,gersh}). It has been
shown\cite{gersh} that the
$1d$ samples are driven into the insulating state by  {\em both} the
WL and interaction effects.

\vspace{0.1cm}

\noindent

 I.L. Aleiner$^{1}$, B.L. Altshuler$^{2,3}$,
and
M.E. Gershenson$^4$\\
$^{1}$SUNY at Stony Brook, Stony Brook, NY 11794\\
$^{2}$Princeton University, Princeton, NJ 08544\\
$^{3}$NEC Research Institute, Princeton, NJ 08540\\
$^{4}$Rutgers University, Piscataway, NJ 08854-8019

\end{multicols}

\end{document}